\documentclass[aps,prl,superscriptaddress,reprint,notitlepage]{revtex4-2}

\usepackage{bm,color}
\usepackage{graphicx}
\usepackage{amsmath, amssymb}
\usepackage{braket}
\usepackage{comment}
\usepackage[compat=1.1.0]{tikz-feynman}
\usepackage{txfonts}

\usepackage[utf8]{inputenc}

\usepackage[hidelinks]{hyperref}

\usepackage{hyperref}
\usepackage{xcolor}
\hypersetup{colorlinks=true}
\usepackage{graphicx}
\usepackage{bm}
\usepackage{amsmath}
\usepackage{amssymb}
\usepackage{xspace}
\usepackage{algorithmic}
\usepackage{algorithm}
\usepackage[capitalise]{cleveref}
\usepackage{txfonts}
\graphicspath{{./figures/}}
\usepackage{physics}
\usepackage{siunitx}
\usepackage{booktabs}
\usepackage{chemformula}
\usepackage{ulem}
\usepackage{orcidlink}
\newcommand{\subfigref}[2]{Fig.~\hyperref[#1]{\ref*{#1}#2}}

\begin{document}
\title{
    Odd-Parity Magnons in the Haldane-Hubbard Model from Topological Exciton Condensation
}
\author{Rintaro Eto\orcidlink{https://orcid.org/0000-0001-8833-1311}}
\email{rintaro.eto@tum.de}
\affiliation{Technical University of Munich, TUM School of Natural Sciences, Physics Department, 85748 Garching, Germany}
\affiliation{Munich Center for Quantum Science and Technology (MCQST), Schellingstraße 4, 80799 München, Germany}
\author{Johannes Knolle\orcidlink{https://orcid.org/0000-0002-0956-2419}}
\affiliation{Technical University of Munich, TUM School of Natural Sciences, Physics Department, 85748 Garching, Germany}
\affiliation{Munich Center for Quantum Science and Technology (MCQST), Schellingstraße 4, 80799 München, Germany}
\date{\today} 
\begin{abstract}
    Odd-wave magnets are the counterparts to even-wave altermagnets  realizing odd-parity spin splitting. Normally discussed for noncollinear systems, they have recently been shown to appear in collinear magnetic states in the presence of loop currents. Here we study collective excitations of the paramagnetic and magnetic phase of the seminal Haldane-Hubbard model. 
    We identify the existence topological excitons in the paramagnetic phase, and their condensation as the driving mechanism into the collinear N\'{e}el state. The latter realizes an odd-wave magnet with odd-parity magnons displaying a characteristic $f$-wave splitting. We further uncover that an electron bandgap closing ensures magnon bandgap closing causing a change in odd-parity magnon topology, as well as a drastically enlarged spin splitting.
    Our results establish the presence of topological excitons and odd-parity magnons in the Haldane-Hubbard, with potential realizations in Floquet-driven materials and cold atomic gases.
\end{abstract}
\maketitle

\begin{figure}[]
    \centering
    \includegraphics[width=0.50\textwidth]{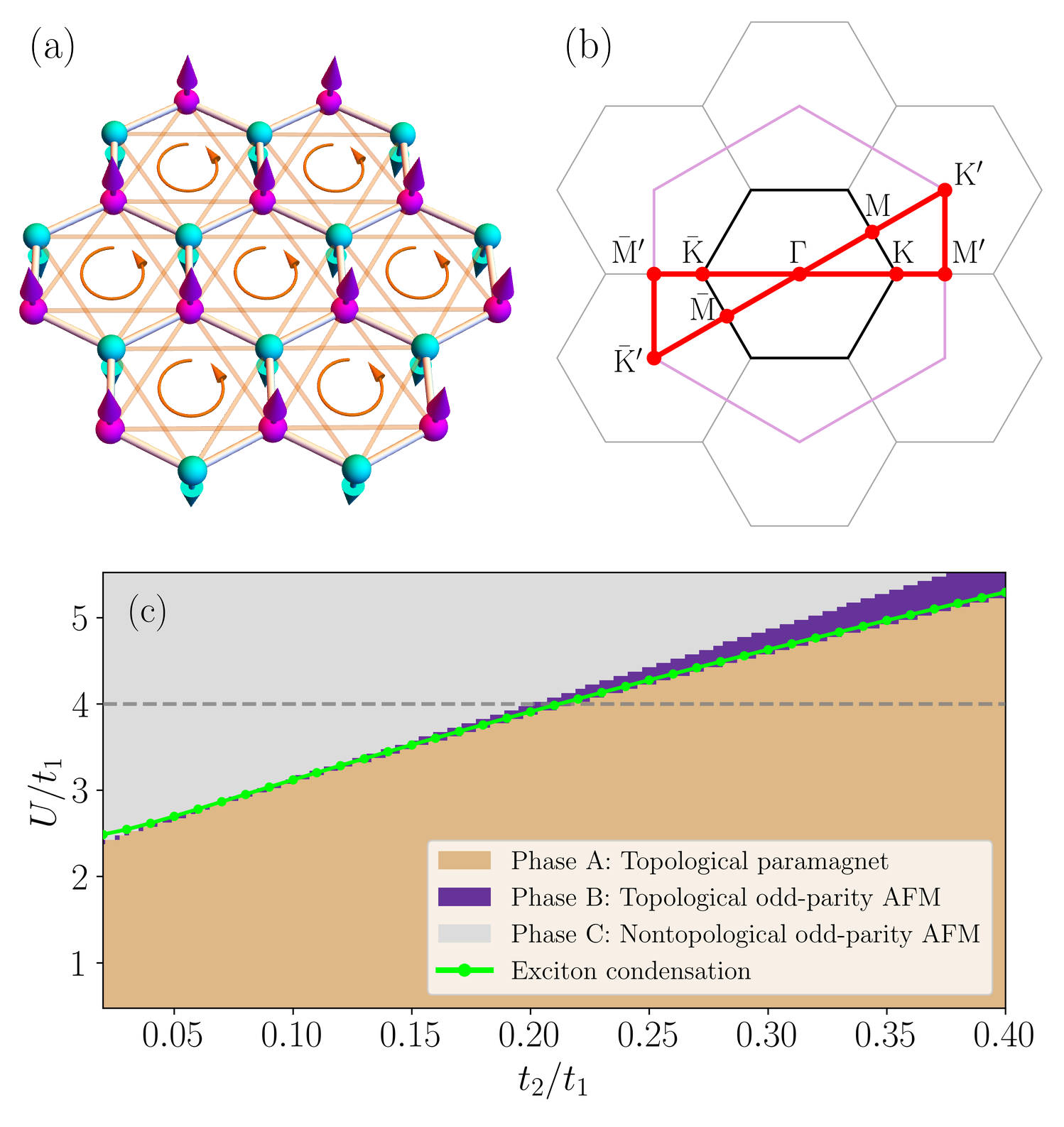}
    \caption{
    (a) Schematic of the 
    odd-parity-wave magnetism in the Haldane-Hubbard (HH) model.
    Bravais vectors are given by $\mathbf{a}_1=(1,0)$, $\mathbf{a}_2=(-1/2,\sqrt{3}/2)$, and $\mathbf{a}_3=(-1/2,-\sqrt{3}/2)$.
    Magenta and cyan arrows represent the collinear N\'{e}el order.
    Orange bonds and arrows denote loop currents encoded by imaginary hoppings.
    (b) Brillouin zones (BZs) of the Haldane-Hubbard model. Black, gray, and magenta hexagons represent the first, second, and an extended BZs, respectively. Red lines and markers describe a certain momentum path and high-symmetry points referred in this work.
    (c) Hartree-Fock phase diagram of the HH model (color map), overlapped by paramagnetic-magnetic phase boundary (green line with markers) obtained with the time-dependent Hartree-Fock approximation.
    Gray dashed line denotes the cut at $U/t_1=4$, along which the dynamical spin structure factors shown in Figs.~\ref{Fig02} and~\ref{Fig04} are evaluated.
    } 
    \label{Fig01}
\end{figure}

Since its proposal in 1988, the Haldane model~\cite{Haldane1988} has served as a paradigmatic example of a Chern (quantum Hall) insulator~\cite{Klitzing1980}, laying the foundation for modern topological band theory and inspiring a wide range of developments, including the Kane-Mele model of quantum spin Hall insulators~\cite{Kane2005Z2QSH,Kane2005QSHGraphene}, and interacting topological phases such as fractional Chern insulators~\cite{Neupert2011}.
More recently, the spinful Haldane model with onsite interactions, known as the Haldane-Hubbard (HH) model, has been extensively studied as a minimal platform for correlated topological matter~\cite{Arun2016,Hickey2016,Vanhala2016,Alba2016,Leite2021,TCYi2023,BQWang2024,WXHe2024,ZLGu2019arXiv}.
Various numerical approaches, including Hartree-Fock~\cite{Arun2016}, exact diagonalization~\cite{Hickey2016}, and dynamical mean-field theory~\cite{Vanhala2016}, have revealed a rich phase diagram featuring diverse magnetic ground states with nontrivial electron topology, such as collinear N\'{e}el and noncoplanar tetrahedral orders~\cite{Arun2016}, and even chiral spin liquids~\cite{Hickey2016}. While the ground state phases of the HH are well established, the nature of the collective excitations has remained largely unexplored. 

The recent establishment of altermagnetism has reshaped our understanding of unconventional magnetic phases by demonstrating that spontaneous spin splitting can emerge \textit{without} relativistic spin-orbit coupling for compensated phases without net magnetization~\cite{LSmejkal2022,LSmejkal2022reviewPRX,IMazin2022editorialPRX}. This development has stimulated broader interest in magnetic phases with nontrivial momentum-space spin-splitting form factors, including odd-parity-wave magnetism. According to the spin-space group classification, odd-wave spin splitting originates from {\it noncollinear} magnetic textures~\cite{Hellenes2023arXiv,Brekke2024,YYu2025,Pari2025,Chakraborty2025,QSong2025,Yamada2025}.
However, recently it has been recognized that analogous odd-parity-wave spin splitting can also arise in {\it collinear} magnets when time-reversal symmetry is additionally broken by loop currents encoded as complex hoppings~\cite{Leeb2026arXiv,YPLin2025arXiv}. In particular, Floquet driving is a well established way of inducing effective loop currents and, thus, collinear odd-parity magnetism induced by circularly polarized light has been proposed in rapid succession as a non-equilibrium realization~\cite{SAAGhorashi2025,PHFu2026,SHuang2026,TZhu2026,BLi2026,TZhang2026,DLiu2026,YTian2026arXiv}. In this context it was realized that the basic N\'{e}el state of the HH realizes an odd-wave magnet with $f$-wave spin splitting~\cite{YPLin2025arXiv}.

The collective excitations of altermagnets, as even-parity-wave magnets, have begun to be actively researched~\cite{Maier2023,Smejkal2023,Costa2025,Garcia-Gaitan2025,YLiu2025,Eto2025}, but those of odd-parity-wave magnets remain largely unexplored. Very recently, Refs.~\cite{Neumann2026arXiv,Kravchuk2026arXiv} studied effective spin models for noncollinear odd-wave magnets but the effects of itinerancy and collinearity have not been addressed.
The Haldane-Hubbard model, thus, provides a natural platform for investigating the interplay between spin density wave N\'{e}el order and electronic loop currents on the magnetic excitations of odd-parity-wave magnets.

In this Letter, we study collective dynamics of the HH model in the paramagnetic and collinear N\'{e}el phases using the time-dependent Hartree–Fock (TDHF) approximation~\cite{Thouless1960,Thouless1961,McLACHLAN1964,Negele1982,Nalabothula2026,TTan2025arXiv} and the self-consistent random phase approximation (RPA). First, starting in the paramagnetic phase, we find sharp topological exciton bands with nonzero Chern numbers inherited from the electronic bands. Second, we show that it is the condensation of these excitons driving the transition from the paramagnetic to the collinear N\'{e}el phase. Third, the resulting magnetically ordered state is an odd-wave magnet with $f$-wave spin splitting not only of electronic bands but also of the resulting magnons.
Fourth, we demonstrate that closing of the electron band gap necessarily accompanies magnon gap closing, leading to a change in the odd-parity magnon topology. Overall, our work uncovers topological exciton condensation as the mechanism for the topological N\'{e}el state of the HH. In addition, we show how excitons transform into odd-parity-wave magnons and that the spin splitting can be greatly enhanced by level repulsion with the particle-hole continuum absent in purely localized spin model descriptions.

\textit{Model.}---The Hamiltonian of the HH model on the two-dimensional honeycomb lattice is given by
\begin{equation}
\begin{aligned}
    \mathcal{H} &= 
    -t_1\sum_{\ev{\mathbf{r},\mathbf{r}'}_\mathrm{NN}} \sum_{\sigma=\uparrow,\downarrow} 
    \left( \hat{c}^\dagger_{\mathbf{r}\sigma}\hat{c}_{\mathbf{r}'\sigma} +\mathrm{h.c.} \right) \\
    &\quad
    -t_2\sum_{\ev{\mathbf{r},\mathbf{r}'}_\mathrm{NNN}} \sum_{\sigma=\uparrow,\downarrow} 
    \left( e^{\mathsf{i}\nu_{\mathbf{r},\mathbf{r}'}\phi} \hat{c}^\dagger_{\mathbf{r}\sigma}\hat{c}_{\mathbf{r}'\sigma} +\mathrm{h.c.} \right) 
    + U\sum_\mathbf{r} \hat{n}_{\mathbf{r}\uparrow} \hat{n}_{\mathbf{r}\downarrow},
    \label{eq:Ham}
\end{aligned}
\end{equation}
where $\hat{c}^\dagger_{\mathbf{r}\sigma}$ $(\hat{c}_{\mathbf{r}\sigma})$ denotes an electron creation (annihilation) operator with spin $\sigma$ at site $\mathbf{r}$. $\hat{n}_{\mathbf{r}\sigma}=\hat{c}^\dagger_{\mathbf{r}\sigma}\hat{c}_{\mathbf{r}\sigma}$ is the number operator.
The first two terms correspond to the Haldane model, i.e., the nearest-neighbor (NN) and the next-nearest-neighbor (NNN) electron hopping on the honeycomb lattice, respectively. Note that $t_1,t_2\geq0$. $\nu_{\mathbf{r},\mathbf{r}'}=\pm1$ induces a spatially uniform flux pattern with a magnitude of $6\phi$ per plaquette, as depicted in Fig.~\ref{Fig01}(a). We set $\phi=\pi/2$ throughout this work, such that the NNN hopping is purely imaginary. The third term is the onsite Hubbard repulsion. We consider the half-filled case and mainly discuss the intermediate interaction regime $U/t_1=4$.
Before studying the dynamical properties of the HH model, we revisit its Hartree-Fock (HF) phase diagram originally presented about a decade ago~\cite{Arun2016}, see Fig.~\ref{Fig01}(c).
When $t_2/t_1 \lesssim 0.4$, the system exhibits three distinct phases:
(A) a topological paramagnetic phase with $C_{v\uparrow} = C_{v\downarrow} = -C_{c\uparrow} = -C_{c\downarrow} = +1$,
(B) a topological collinear antiferromagnetic (AFM) phase in which the electron bands have nontrivial Chern numbers of $C_{v\uparrow} = C_{v\downarrow} = -C_{c\uparrow} = -C_{c\downarrow} = +1$, and
(C) a nontopological collinear AFM phase,
$c$ and $v$ are unoccupied and occupied band indices, respectively.

\textit{Time-Dependent Hartree-Fock}.---Throughout this work, we study particle-hole excitations in the HH model using the TDHF approximation. In TDHF, exciton eigenmodes are obtained with the eigenvalue problem~\cite{Hirsbrunner2026arXiv}
\begin{equation}
    \sum_{\sigma\bar{\sigma}} \left\{
    \left[ \mathcal{H}, Q^\dagger_{\mathbf{q},\nu,\sigma\rightarrow \bar{\sigma}} \right]
    -
    \hbar\omega_{\mathbf{q},\nu} Q^\dagger_{\mathbf{q},\nu,\sigma\rightarrow \bar{\sigma}}
    \right\} \ket{\mathsf{0}} = 0
\end{equation}
where the creation operator of an exciton as a spin-flip $(\sigma\rightarrow\bar{\sigma})$ particle-hole pair in the Bogoliubov-de-Gennes (BdG) form
$Q^\dagger_{\mathbf{q},\nu,\sigma\rightarrow \bar{\sigma}}$
is given by
\begin{equation}
\begin{aligned}
    Q^\dagger_{\mathbf{q},\nu,\sigma\rightarrow \bar{\sigma}}
    &= \sum_\mathbf{k} \left\{ 
    X_{\mathbf{q},\nu}^{\sigma\rightarrow \bar{\sigma}}(\mathbf{k}-(1-\alpha)\mathbf{q}) \hat{c}^\dagger_{\mathbf{k}+\alpha\mathbf{q},c\bar{\sigma}}\hat{c}_{\mathbf{k}-(1-\alpha)\mathbf{q},v\sigma} \right. \\
    &\quad\quad - \left.
    Y_{\mathbf{q},\nu}^{\sigma\rightarrow \bar{\sigma}}(\mathbf{k}-(1-\alpha)\mathbf{q}) \hat{c}^\dagger_{-\mathbf{k}+(1-\alpha)\mathbf{q},v\bar{\sigma}}\hat{c}_{-\mathbf{k}-\alpha\mathbf{q},c\sigma}
    \right\},
\end{aligned}
\end{equation}
where
$X_{\mathbf{q},\nu}^{\sigma\rightarrow \bar{\sigma}}(\mathbf{k}-(1-\alpha)\mathbf{q})$ and
$Y_{\mathbf{q},\nu}^{\sigma\rightarrow \bar{\sigma}}(\mathbf{k}-(1-\alpha)\mathbf{q})$ are the envelope functions of excitons.
Note that $\nu$ is an index of exciton bands.
$\ket{\mathsf{0}}$ denotes the correlated ground state.
Different choices of Berry connection correspond to different conventions for assigning a collective position coordinate to the particle-hole excitation. These conventions can be parameterized by $\alpha\in[0,1]$~\cite{YHKwan2021,Davenport2026ExcitonBerryology,Davenport2026rhombohedralgraphene}. We confirmed that $\alpha=0$ and $\alpha=1$ yield identical results up to negligible numerical error~\cite{Hirsbrunner2026arXiv}. We therefore present the results with $\alpha=1$ hereafter.

For finite size systems with periodic boundary condition, TDHF calculation for the downward spin-flip $(\uparrow\rightarrow\downarrow)$ sector of the HH model is essentially given by the following paraunitary eigenvalue problem~\cite{WTZhou2026,Hirsbrunner2026arXiv}:
\begin{equation}
\begin{aligned}
    &\sum_\mathbf{k}
    \left(
    \begin{array}{cc}
          A_\mathbf{q}^{\uparrow\rightarrow \downarrow}(\mathbf{k}',\mathbf{k}) 
        & B_\mathbf{q}^{\uparrow\rightarrow \downarrow}(\mathbf{k}',\mathbf{k}) \\
          \bar{B}_{\mathbf{q}}^{\uparrow\rightarrow \downarrow}(\mathbf{k}',\mathbf{k})
        & \bar{A}_{\mathbf{q}}^{\uparrow\rightarrow \downarrow}(\mathbf{k}',\mathbf{k})
    \end{array}
    \right)
    \left(
    \begin{array}{c}
        X_{\mathbf{q},\nu}^{\uparrow\rightarrow \downarrow}(\mathbf{k}) \\
        Y_{\mathbf{q},\nu}^{\uparrow\rightarrow \downarrow}(\mathbf{k}) 
    \end{array}
    \right)
    \\ &\quad\quad = 
    \omega_{\mathbf{q},+-}
    \left(
    \begin{array}{cc}
        \mathsf{1} & \mathsf{0} \\ \mathsf{0} & -\mathsf{1}
    \end{array}
    \right)
    \left(
    \begin{array}{c}
        X_{\mathbf{q},\nu}^{\uparrow\rightarrow \downarrow}(\mathbf{k}') \\
        Y_{\mathbf{q},\nu}^{\uparrow\rightarrow \downarrow}(\mathbf{k}')
    \end{array}
    \right),
\end{aligned}
\end{equation}
where
\begin{equation}
\begin{aligned}
    &A_\mathbf{q}^{\uparrow\rightarrow \downarrow}(\mathbf{k}',\mathbf{k})
    = \left( \varepsilon_{\mathbf{k}+\mathbf{q},c\downarrow} - \varepsilon_{\mathbf{k},v\uparrow} \right) \delta_{\mathbf{k},\mathbf{k}'} \\
    &\quad - \frac{U}{N} \sum_\alpha 
    \ev{\mathbf{k} ,v|\sigma^+_\alpha|\mathbf{k} +\mathbf{q},c}
    \ev{\mathbf{k}'+\mathbf{q},c|\sigma^-_\alpha|\mathbf{k}',v},
\end{aligned}
\end{equation}
\begin{equation}
\begin{aligned}
    &B_\mathbf{q}^{\uparrow\rightarrow \downarrow}(\mathbf{k}',\mathbf{k}) \\
    &\quad = +\frac{U}{N} \sum_\alpha 
    \ev{-\mathbf{k}-\mathbf{q},c|\sigma^+_\alpha|-\mathbf{k},v}
    \ev{\mathbf{k}'+\mathbf{q},c|\sigma^-_\alpha|\mathbf{k}',v},
\end{aligned}
\end{equation}
\begin{equation}
\begin{aligned}
    &\bar{B}_\mathbf{q}^{\uparrow\rightarrow \downarrow}(\mathbf{k}',\mathbf{k}) \\
    &\quad = +\frac{U}{N} \sum_\alpha 
    \ev{\mathbf{k} ,v|\sigma^+_\alpha|\mathbf{k} +\mathbf{q},c}
    \ev{-\mathbf{k}',v|\sigma^-_\alpha|-\mathbf{k}'-\mathbf{q},c},
\end{aligned}
\end{equation}
and
\begin{equation}
\begin{aligned}
    &\bar{A}_\mathbf{q}^{\uparrow\rightarrow \downarrow}(\mathbf{k}',\mathbf{k})
    = \left( \varepsilon_{-\mathbf{k}-\mathbf{q},c\uparrow} - \varepsilon_{-\mathbf{k},v\downarrow} \right) \delta_{\mathbf{k},\mathbf{k}'} \\
    &\quad - \frac{U}{N} \sum_\alpha 
    \ev{-\mathbf{k} -\mathbf{q},c|\sigma^+_\alpha|-\mathbf{k} ,v}
    \ev{-\mathbf{k}',v|\sigma^-_\alpha|-\mathbf{k}'-\mathbf{q},c}.
\end{aligned}
\end{equation}
$\sigma^\pm_\alpha=(1/2)(\sigma^x_\alpha\pm\mathsf{i}\sigma^y_\alpha)$ denote the Pauli spin-flip operators acting on the sublattice-$\alpha$ $(=\mathrm{A},\mathrm{B})$ basis.
$\ket{\mathbf{k},\nu}$ is a Bloch state obtained within the self-consistent Hartree-Fock approximation [See End Matter] with momentum $\mathbf{k}$ and a band index $\nu$.
We consider system sizes up to $30^2$ unit cells and find that finite size effects seen in the dispersion relation are negligibly small. We solve the paraunitary eigenvalue problem using Cholesky decomposition followed by a usual unitary diagonalization~\cite{Colpa1978}. Chern numbers of the lowest exciton modes in each sector are calculated using the Wilson loop method~\cite{Fukui2005,Hirsbrunner2026arXiv} respecting $D_{6}$ symmetric distribution of the $\mathbf{k}$-mesh.

A key advantage of TDHF is that it provides a unified description of collective excitations across the excitonic condensation transition. Within the same formalism, excitons in the paramagnetic phase continuously evolve into magnons in the magnetically ordered phase, enabling a direct characterization of the change in quasiparticle nature associated with symmetry breaking~\cite{Olsen2021,WTZhou2026}.

\begin{figure}
    \centering
    \includegraphics[width=0.50\textwidth]{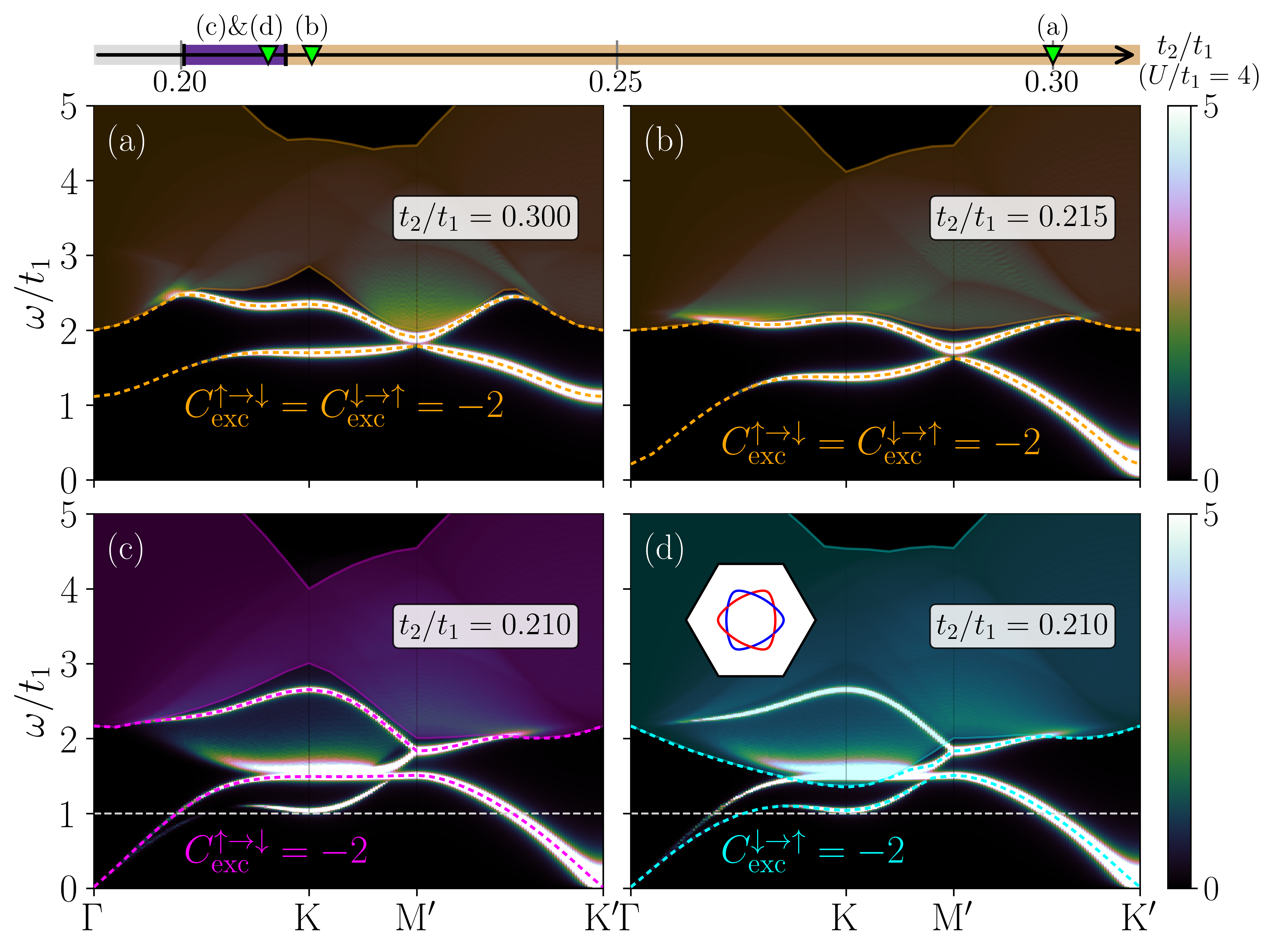}
    \caption{
    (a),(b) Dynamical spin structure factor
    $S(\mathbf{q},\omega)=S_{+-}(\mathbf{q},\omega)+S_{-+}(\mathbf{q},\omega)$
    of the HH model, overlaid with the TDHF spectra in the topological paramagnetic phase A.
    Colored dashed lines and shaded regions represent isolated collective modes and
    particle-hole continua, respectively.
    (c),(d) Same as (a),(b), but for the topological odd-parity-wave AFM phase B,
    together with the spin-resolved TDHF spectra.
    The inset in (d) shows the isoenergy contour of the TDHF spectrum at
    $\omega/t_1=1$, corresponding to the white dashed lines in (c) and (d).
    For all panels, $U/t_1=4$ and $\phi=\pi/2$.
    The values of $t_2/t_1$ are indicated in each panel and marked in the phase
    diagram at the top.
    Light gray, purple, and orange bands in the phase diagram show phases C, B, and A, respectively.
    }
    \label{Fig02}
\end{figure}

\begin{figure}
    \centering
    \includegraphics[width=0.40\textwidth]{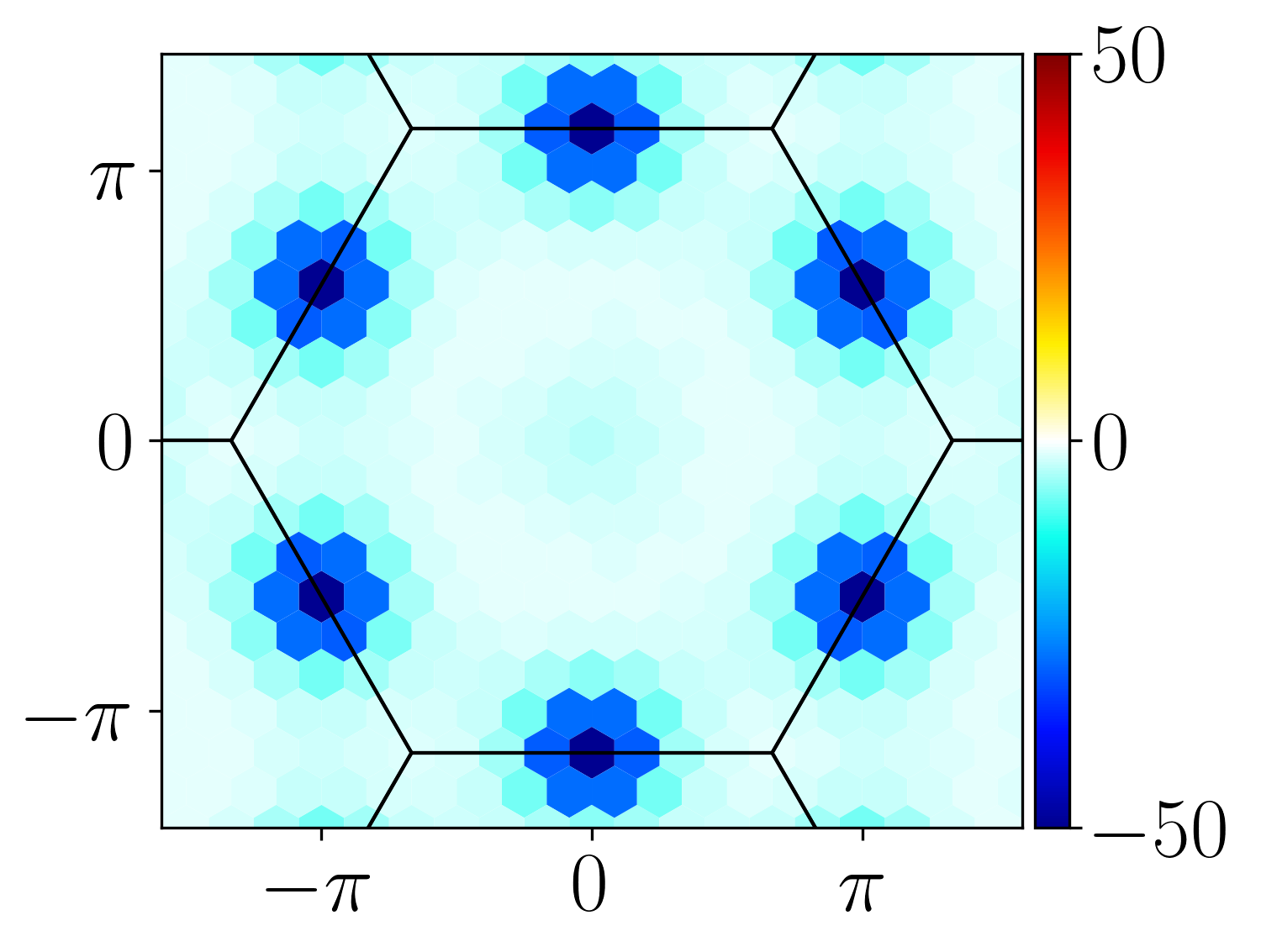}
    \caption{
    Exciton Berry curvature of the lowest bands with $C_\textrm{exc}^{\uparrow\rightarrow\downarrow(\downarrow\rightarrow\uparrow)}=-2$ in the paramagnetic phase A at $t_2/t_1=0.3$, $U/t_1=4$, and $\phi=\pi/2$. Note that the Berry curvature distribution is identical for $+-(\uparrow\rightarrow\downarrow)$ and $-+(\downarrow\rightarrow\uparrow)$ excitons. Black hexagon denotes the first Brillouin zone.
    } 
    \label{Fig03}
\end{figure}

\begin{figure*}
    \centering
    \includegraphics[width=\textwidth]{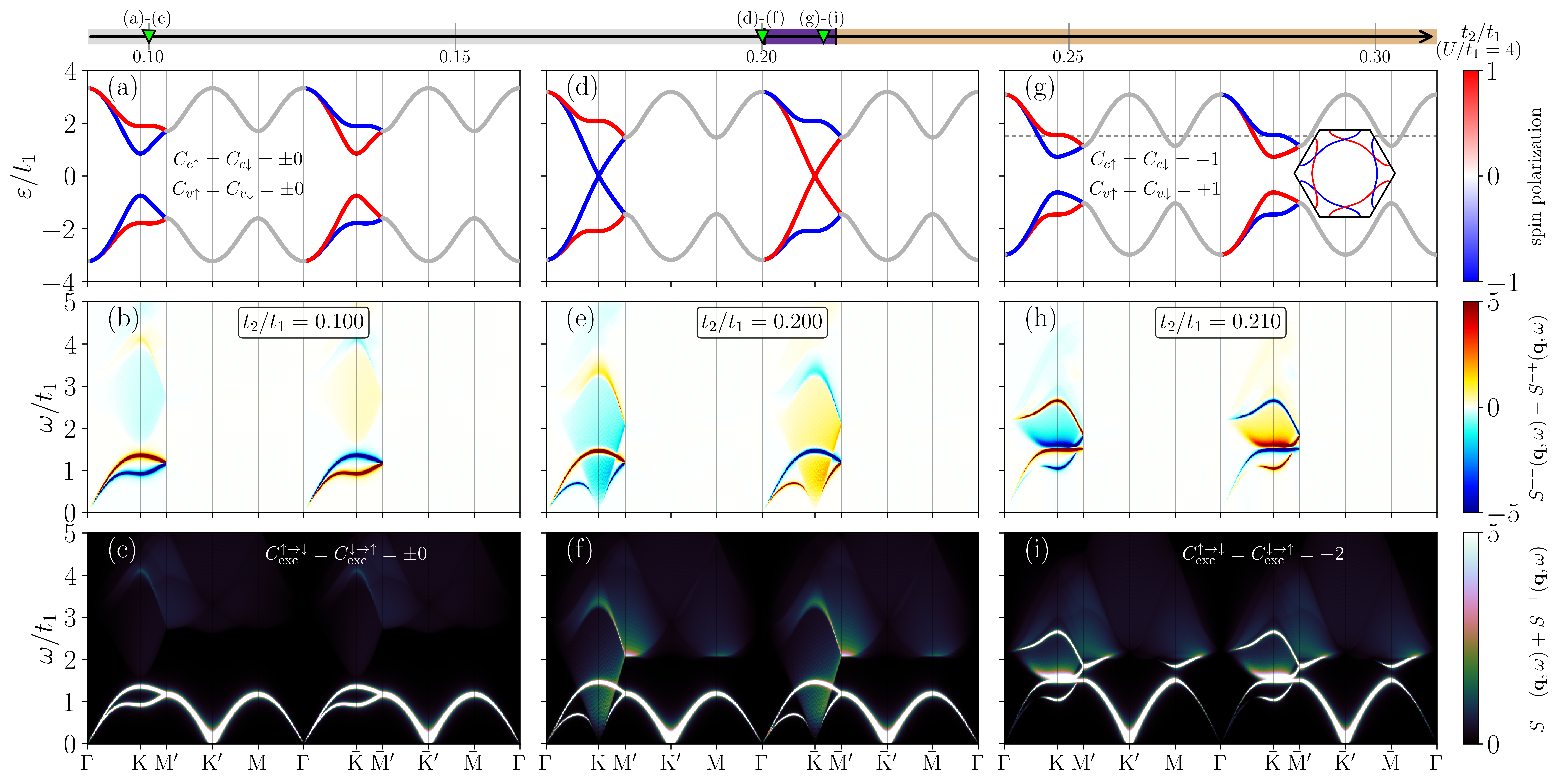}
    \caption{
    (a),(d),(g) Electron bands near the nontopological-topological AFM phase boundary obtained with Hartree-Fock approximation. Red, blue and gray bands indicate up-spin, down-spin, and spin-degenerate ones, respectively.
    The inset in (g) shows the isoenergy contour of the electron bands at
    $\varepsilon/t_1=1.5$, corresponding to the gray dashed line.
    (b),(e),(h) Difference in spin-resolved dynamical spin-structure factors $S_{+-}(\mathbf{q},\omega)-S_{-+}(\mathbf{q},\omega)$.
    (c),(f),(i) Dynamical spin-structure factor
    $S(\mathbf{q},\omega)=S_{+-}(\mathbf{q},\omega)+S_{-+}(\mathbf{q},\omega)$.
    For all panels, $U/t_1=4$ and $\phi=\pi/2$. The value of $t_2/t_1$ is common to each column and is indicated in the corresponding panels as well as marked in the phase diagram at the top.
    Light gray, purple, and orange bands in the phase diagram show phases C, B, and A, respectively.
    } 
    \label{Fig04}
\end{figure*}

\textit{Topological Excitons}.---First, we concentrate on the paramagnetic phase and show in Fig.~\ref{Fig02}(a) the dynamical spin-structure factor $S(\mathbf{q},\omega)$ obtained with the RPA, see End Matter for details, along the high-symmetry momentum path given in Fig.~\ref{Fig01}(b) which is overlayed with the TDHF exciton bands $\left\{\omega_{\mathbf{q},\nu}\right\}$. Both RPA and TDHF show the  existence of sharp low-energy modes, i.e. excitons. The doubly-degenerate lowest exciton bands with opposite spin-flip processes $\nu=+-(\uparrow\rightarrow\downarrow)$ and $\nu=-+(\downarrow\rightarrow\uparrow)$, are decoupled from each other, are isolated from the upper particle-hole continuum by a tiny gap at the M point, and each possesses a Chern number of $C_\textrm{exc}^{\uparrow\rightarrow\downarrow(\downarrow\rightarrow\uparrow)}=-2$. The exciton Chern number can be understood to originate from the difference in Chern numbers between unoccupied and occupied electron bands~\cite{HYXie2024}, i.e.,
\begin{equation}
      C_\mathrm{exc}^{\sigma\rightarrow\bar{\sigma}} 
    = C_{c\bar{\sigma}} - C_{v\sigma}.
    \label{eq:relation_exciton_electron_Chern_number}
\end{equation}
Thus, we have revealed sharp topological excitons as novel collective excitations already at intermediate interaction strength in the paradigmatic HH model.

Figure~\ref{Fig03} shows the Berry curvature (BC) distribution of the lowest exciton band, which is identical for the upward and downward spin-flip sectors, $\uparrow\rightarrow\downarrow$ and $\downarrow\rightarrow\uparrow$. Sharp BC peaks appear at each of the three inequivalent M points, where the gap between the lowest exciton band and the upper continuum is minimal [see Fig.~\ref{Fig02}(a)]. This exciton BC distribution is in stark contrast to that of the underlying electron bands, whose nontrivial topology originates from band-gap closing and reopening at the K and $\bar{\mathrm{K}}$ points and is therefore accompanied by BC accumulation around them. Remarkably, despite the simple relation $C_{\mathrm{exc}}^{\sigma\rightarrow\bar{\sigma}}=C_{c\bar{\sigma}}-C_{v\sigma}$, the exciton BC is concentrated near the M points rather than the K and $\bar{\mathrm{K}}$ points that govern the electronic topology. This observation highlights that the momentum-space structure of exciton topology is not trivially inherited from that of the constituent electron and hole bands, but is instead strongly influenced by the collective nature of the particle-hole bound state.
Notably, the exciton BC remains finite even near the optically accessible $\Gamma$ point owing to broken time-reversal symmetry. This finite long-wavelength BC can generate an anomalous transverse velocity of exciton wave packets under an external potential or strain gradient, potentially leading to an exciton Hall drift.



Exciton topology is sometimes led by underlying crystalline symmetries~\cite{Davenport2024}.
It is noteworthy that the nontrivial exciton Chern number $C_\textrm{exc}^{\uparrow\rightarrow\downarrow(\downarrow\rightarrow\uparrow)}=\pm2\mod6$ here can be understood from recently proposed stable exciton wavefunction (EWF) zeros~\cite{YHwang2026arXiv}. We extend this concept to BdG systems, in which the EWF intensity is given by
\begin{equation}
    \mathsf{m}_{\mathbf{q},\nu}^{\sigma\rightarrow \bar{\sigma}}(\mathbf{k}) =
    \left| X_{\mathbf{q},\nu}^{\sigma\rightarrow \bar{\sigma}}(\mathbf{k}) \right|^2 -
    \left| Y_{\mathbf{q},\nu}^{\sigma\rightarrow \bar{\sigma}}(\mathbf{k}) \right|^2.
\end{equation}
In the topological paramagnetic phase of the HH model, it is numerically found that
$\mathsf{m}_{\mathbf{q}=\Gamma,\nu=\mathrm{lowest}}^{\sigma\rightarrow \bar{\sigma}}(\mathrm{K})=\mathsf{m}_{\mathbf{q}=\Gamma,\nu=\mathrm{lowest}}^{\sigma\rightarrow \bar{\sigma}}(\bar{\mathrm{K}})=0$ and $\mathsf{m}_{\mathbf{q}=\Gamma,\nu=\mathrm{lowest}}^{\sigma\rightarrow \bar{\sigma}}(\mathbf{k})\neq0$ for arbitrary $\mathbf{k}\in\{\mathbf{k}\}_{\in\mathrm{BZ}}\setminus(\mathrm{K}\lor\bar{\mathrm{K}})$. This EWF zero distribution is consistent with a constraint $C_\textrm{exc}=\pm2\mod6$ in $C_6$ symmetric systems~\cite{YHwang2026arXiv}.

\textit{Exciton Condensation and $f$-Wave Spin-Split Magnons}---Next, we discuss how the topological excitons behave in the vicinity of the phase boundary between the paramagnetic and magnetic phases. Figure~\ref{Fig02}(b) shows the excitation spectrum when $(t_2,U)/t_1=(0.215,4)$, which is very close to the phase boundary yet still in the paramagnetic phase. The topological excitons are ready to condensate at the $\Gamma$ point. Their condensation at $(t_2,U)/t_1\approx(0.212,4)$ drives the phase transition into the topological AFM phase B. It is also worth noting that, as indicated in Fig.~\ref{Fig01}(c), the exciton condensation obtained with the TDHF almost perfectly coincides with the nonmagnetic-magnetic phase boundary obtained with HF, suggesting that the mean-field transition can be understood in terms of topological exciton condensation.

The condensed excitons transform into magnons, i.e. exhibit linearly dispersive Type-I Goldstone mode behavior. Remarkably, the magnon modes of the odd-parity magnetic state display $f$-wave spin splitting, as indicated in Figs.~\ref{Fig02}(c), \ref{Fig02}(d) and its inset. The dispersion relation of magnons with upward and downward spin-flip processes are nondegenerate at general wavenumbers and are connected by
$\omega_{\mathbf{q},\pm\mp} = C_{3\parallel}^{\pm1}\omega_{-\mathbf{q},\mp\pm}$
where $C_{3\parallel}$ denotes threefold rotation around the out-of-plane axis.
Note that the long-wavelength behavior of the odd-parity-wave magnons as Type-I Goldstone modes is ensured by the spontaneous breaking of the spin-space SO(3) symmetry into SO(2).
Our calculation present the first identification of nonrelativistic collinear odd-parity-wave magnons in an itinerant model.
Note, the $f$-wave magnon spin-splitting is associated with those of electron bands induced by loop currents~\cite{TZhu2026,Leeb2026arXiv,YPLin2025arXiv}, where spin-space $C_{2\perp}\mathcal{T}$ is broken while spin-space $C_{2\perp}$ followed by real-space inversion $\mathcal{P}$ is preserved. The Chern numbers of spin-split magnons are $-2$ for both upward and downward spin flips as shown in Figs.~\ref{Fig02}(c) and \ref{Fig02}(d), indicating that Eq.~(\ref{eq:relation_exciton_electron_Chern_number}) connecting exciton and electron Chern numbers remains valid even in the topological odd-parity-wave AFM phase B.

\textit{Magnon Bandgap Closing Enforced by Electron Bandgap Closing}.---We next discuss the transition from a topological to a nontopological AFM phase.
It is long known that the phase transition originates from the gap closing and reopening of the $f$-wave spin-split electron bands~\cite{Arun2016,TZhu2026} at K and $\bar{\mathrm{K}}$ points, which is reproduced in Figs.~\ref{Fig04}(a),~\ref{Fig04}(d), and~\ref{Fig04}(g). This gap closing induces a downward shift of the spin-resolved particle-hole continuum towards low energy  at K and $\bar{\mathrm{K}}$ points, as clearly seen in Figs.~\ref{Fig04}(c),~\ref{Fig04}(f), and~\ref{Fig04}(i). In particular, when the electron bandgap closes for $(t_2,U)/t_1\approx(0.200.4)$, the continuum touches zero energy, as shown in Fig.~\ref{Fig04}(f). Since $f$-wave spin-split magnons are excitonic particle-hole bound states, thus always lower in energy than the particle-hole continuum, it is ensured that such magnons touch the upper continua. This \textit{enforced} magnon bandgap closing leads to a change in magnon topology from $C_\mathrm{exc}^{\bar{\sigma}\rightarrow\sigma}=-2$ to $C_\mathrm{exc}^{\bar{\sigma}\rightarrow\sigma}=\pm0$, highlighting nontrivial correlation between magnons as particle-hole bound states and the upper particle-hole continua~\cite{XCai2026}. This also ensures the validity of Eq.~(\ref{eq:relation_exciton_electron_Chern_number}) even in the nontopological AFM phase A.

It is also worth pointing out that, within the topological odd-parity-wave AFM phase B, spin-split magnons exhibit \textit{avoided decay}-like features~\cite{Verresen2019,XWang2024} near K and $\bar{\mathrm{K}}$ points, in which bound states are pushed out of the continuum by strong interaction effects, as seen in Figs.~\ref{Fig04}(e) and~\ref{Fig04}(h). As a result, the magnitude of $f$-wave magnon spin-splitting is drastically enhanced. Note that a similar enhanced spin-splitting by the avoided decay-like mechanism has recently been reported for even-parity altermagnets near the metal-insulator transition~\cite{Issing2026arXiv}.

\textit{Discussion}.---We have provided a comprehensive study of collective magnetic excitations in the Haldane-Hubbard model. We identified sharp topological exciton modes in the paramagnetic phase and showed how their condensation induces an $f$-wave odd-parity collinear magnetic phase. We then computed the spin-split odd-parity magnons and their topological properties in relation to the underlying electron band topology. Our work provides the first identification of odd-parity-wave magnons in nonrelativistic collinear systems.

Our findings can be, in principle, experimentally tested in Floquet-engineered collinear antiferromagnetic materials~\cite{SAAGhorashi2025,PHFu2026,SHuang2026,TZhu2026,BLi2026,TZhang2026,DLiu2026,YTian2026arXiv}.
We not only expect odd-parity-wave spin-splitting of electrons and magnons in the magnetically ordered phases but also measurable collective exciton modes in the paramagnetic phase, e.g. even when heating effect destroys long range magnetic order.
Moreover, even paramagnetic materials such as graphene~\cite{Oka2009,Sentef2015,Schueler2020,Merboldt2025,Sentef2026} can be targeted to observe topological excitons.
Apart from solid state materials we also expected that some of our findings could be realized in cold-atom systems, in which the original noninteracting Haldane model has already been realized~\cite{Jotzu2014,Aidelsburger2015}. Understanding non-equilibrium observables beyond linear response will be crucial for the identification of odd-parity magnetism in such quantum simulator platforms.

\textit{Note added}.---Upon completion of this work, we became aware of Ref.~\cite{PZhang2026arXiv}, which studies Floquet nonrelativistic odd-parity-wave magnons using an effective spin model.

\textit{Acknowledgement}.---We thank Henry Davenport for helpful feedback to this manuscript. R.E. thanks Alexander Mook for pointing out Ref.~\cite{WTZhou2026}. J.K. thanks Valentin Leeb for discussion and related collaborations. We acknowledge support from the Deutsche Forschungsgemeinschaft (DFG, German Research Foundation) under Germany’s Excellence Strategy (EXC–2111–390814868 and ct.qmat EXC-2147-390858490), and DFG Grants No. KN1254/1-2, KN1254/2-1 TRR 360 - 492547816 and SFB 1143 (project-id 247310070), as well as the Munich Quantum Valley, which is supported by the Bavarian state government with funds from the Hightech Agenda Bayern Plus. JK thanks the Keck foundation for support.
R.E. acknowledges financial support by JSPS Overseas Fellowship.

\clearpage

\section{End Matter}

\subsection{Hartree-Fock}

In the Hartree-Fock approximation, we decompose the onsite Hubbard interaction in Eq.~(\ref{eq:Ham}) as follows:
\begin{equation}
    \hat{n}_{\mathbf{r}\uparrow} \hat{n}_{\mathbf{r}\downarrow}
    \approx
    \langle \hat{n}_{\mathbf{r}\uparrow} \rangle \hat{n}_{\mathbf{r}\downarrow}
    + \hat{n}_{\mathbf{r}\uparrow} \langle \hat{n}_{\mathbf{r}\downarrow} \rangle
    - \langle \hat{n}_{\mathbf{r}\uparrow} \rangle \langle \hat{n}_{\mathbf{r}\downarrow} \rangle.
\end{equation}
Note that, since we are now interested only in paramagnetic and collinear orders, we practically omit the Fock terms with spin flip in our calculation. Indeed, in the parameter region shown in Fig.~\ref{Fig01}(c), noncollinear and noncoplanar orders are not the mean-field ground state~\cite{Arun2016}.

Within the Hartree-Fock approximation, electron bands and Bloch states are obtained with the following momentum-resolved eigenvalue problem
\begin{equation}
    \hat{H}_\sigma^\textrm{MF}(\mathbf{k}) \ket{\mathbf{k},v/c \ \sigma} = \varepsilon_{\mathbf{k},v/c \ \sigma} \ket{\mathbf{k},v/c \ \sigma},
\end{equation}
where
\begin{equation}
    \hat{H}_\sigma^\textrm{MF}(\mathbf{k}) = \hat{\mathbf{c}}^\dagger_\sigma
    \left(
    \begin{array}{cc}
         +\iota(\mathbf{k}) + U\langle \hat{n}_{\mathrm{A}\bar{\sigma}} \rangle & \gamma(\mathbf{k}) \\
         \gamma^*(\mathbf{k}) & -\iota(\mathbf{k}) + U\langle \hat{n}_{\mathrm{B}\bar{\sigma}} \rangle
    \end{array}
    \right)
    \hat{\mathbf{c}}_\sigma.
\end{equation}
Now $\hat{\mathbf{c}}^\dagger_\sigma=\left( \hat{c}^\dagger_{\mathbf{k},\mathrm{A}\sigma}, \hat{c}^\dagger_{\mathbf{k},\mathrm{B}\sigma}\right)$ and
\begin{align}
    \iota(\mathbf{k}) &= -2t_2 \sum_{i=1,2,3} \cos\left( \mathbf{k}\cdot\mathbf{a}_i+\phi \right), \\
    \gamma(\mathbf{k}) &= -t_1 \left( 1+e^{\mathsf{i}\mathbf{k}\cdot\mathbf{a}_3}+e^{-\mathsf{i}\mathbf{k}\cdot\mathbf{a}_2} \right).
\end{align}
A and B are sublattice indices.

\subsection{Random-Phase Approximation}

Within the RPA, spin-resolved dynamical structure factors are obtained from the RPA susceptibility tensor as
\begin{equation}
    S_{\pm\mp}(\mathbf{q},\omega) = -\frac{1}{\pi}\sum_{\alpha\beta} \mathrm{Im}
    \left\{ \left[ \chi^\mathrm{RPA}_{\pm\mp}(\mathbf{q},\omega+\mathsf{i}\mathsf{0}^+) \right]_{\alpha\beta}
    e^{-\mathsf{i}\mathbf{q}\cdot(\bm{\delta}_\alpha-\bm{\delta}_\beta)}\right\},
\end{equation}
where $\bm{\delta}_\mathrm{A}=(0,0)$ and $\bm{\delta}_\mathrm{B}=(0,1/\sqrt{3})$ denote internal coordinates of sublattices within a unit cell. $\mathsf{0}^+$ denotes an infinitesimal positive number.
The RPA susceptibility tensor is obtained by resumming the bare susceptibility tensor as
\begin{equation}
    \chi^\mathrm{RPA}_{\pm\mp}(\mathbf{q},\mathsf{i}\omega_n)
    = \frac{\chi^0_{\pm\mp}(\mathbf{q},\mathsf{i}\omega_n)}
    {\mathsf{1}-U\chi^0_{\pm\mp}(\mathbf{q},\mathsf{i}\omega_n)},
\end{equation}
where $\mathsf{1}$ denotes an identity.
This formula is valid for arbitrary onsite Hubbard models with spin-$\sigma_z$ conservation and the HH model given by Eq.~(\ref{eq:Ham}) falls into this category.
The explicit form of the bare susceptibility tensor $\chi^0$ is constructed from the Hartree-Fock eigenenergies and eigenstates~\cite{Schrieffer1988,Schrieffer1989,Knolle2010,Knolle2011,Willsher2023} as
\begin{equation}
\begin{aligned}
    \left[\chi^0_{\pm\mp}(\mathbf{q},\mathsf{i}\omega_n) \right]_{\alpha\beta}
    &= -\frac{1}{N_\mathbf{k}} \sum_\mathbf{k} \sum_{\nu_1,\nu_2}
    \frac{f(\varepsilon_{\mathbf{k},\nu_1}-\mu) - f(\varepsilon_{\mathbf{k}+\mathbf{q},\nu_2}-\mu)}{\mathsf{i}\omega_n+\varepsilon_{\mathbf{k},\nu_1}-\varepsilon_{\mathbf{k}+\mathbf{q},\nu_2}} \\
    &\quad\quad \times
    \ev{\mathbf{k},\nu_1|\sigma^{\pm}_\alpha|\mathbf{k}+\mathbf{q},\nu_2}
    \ev{\mathbf{k}+\mathbf{q},\nu_2|\sigma^{\mp}_\beta |\mathbf{k},\nu_1},
\end{aligned}
\end{equation}
where $\omega_n=(2n+1)\pi/k_\textrm{B}T$ denotes the fermionic Matsubara frequency. $f(\varepsilon)=\left[\exp(\varepsilon/k_\textrm{B}T)+1 \right]^{-1}$ denotes the Fermi-Dirac distribution function. $\mu$ is chemical potential.

\bibliography{bib}

\end{document}